\documentclass{osa-article}

\journal{oe}
\begin{document}

\title{Entanglement optimization of filtered output fields in cavity optomechanics}

\author{Xiao-Bo Yan,\authormark{1,*} Zhi-Jiao Deng,\authormark{2} Xue-Dong Tian,\authormark{3} and Jin-Hui Wu\authormark{4,5}}

\address{\authormark{1}College of Electronic Science, Northeast Petroleum University, Daqing 163318, China\\
\authormark{2}Department of Physics, National University of Defense Technology, Changsha 410073, China\\
\authormark{3}College of Physics Science and Technology, Guangxi Normal University, Guilin 541004, China\\
\authormark{4}School of Physics, Northeast Normal University, Changchun 130024, China\\
\authormark{5}jhwu@nenu.edu.cn}

\email{\authormark{*}xiaoboyan@126.com} %% email address is required



\begin{abstract}
Output entanglement is a key element in quantum information processing. Here, we
show how to obtain optimal entanglement between two filtered output
fields in a three-mode optomechanical system. First, we obtain the key analytical expression of optimal time delay between the two
filtered output fields, from which we can obtain the optimal coupling for output entanglement without time delay. In this case, our linearized analysis predicts that the entanglement saturates to an optimal value as the optomechanical coupling
is increased. Furthermore, we obtain the optimal output entanglement with time delay. These
results should be very helpful in conceiving new optomechanical schemes of quantum information
processing with their efficiency depending critically on the degree of output entanglement.
\end{abstract}

%%%%%%%%%%%%%%%%%%%%%%%%%%  body  %%%%%%%%%%%%%%%%%%%%%%%%%%
\section{Introduction}
Entanglement is the distinguishing feature
of quantum mechanics because it is responsible
for nonlocal correlations between observables and now it has become a basic
resource for many quantum information processing
schemes \cite{Samuel2005,Weedbrook2012}. For example, entanglement is required in quantum teleportation in which quantum information can be transmitted from one location to another with security. Extending entanglement into macroscopic systems has become a prominent
experimental objective, and would allow us to explore the quantum-classical
boundary. So far, a number of theoretical and experimental works on entanglement between macroscopic objects
have been studied, such as between atomic ensembles \cite{Julsgaard2011,Krauter2011}, between superconducting
qubits \cite{Berkley2003,Neeley2010,DiCarlo2010,Flurin2012}, and between mechanical
oscillator and microwave fields \cite{Palomaki2013}. Recently, quantum
entanglement in cavity optomechanics has received increasing attention for the
potential to use radiation pressure to generate various entanglement between
subsystems \cite{Mancini2003,Pirandola2003,Pirandola2004,Pirandola2006,Kiesewetter2014,Bhattacharya2008,Chen2014,Liao2014,Yang2015,Paternostro2007,Wipf2008,Genes2008,Barzanjeh2011,Barzanjeh2012,Barzanjeh2013,Wang2013,Kuzyk2013,Vitali2007,Hofer2011,Akram2012,Sinha2015,JLi2015,JLi2017,Asjad2016,Bing2014,Tian2013,Deng2015,Deng2016,Wang2015,Yan2017,Li2013,Sun2017}. Due to the small mechanical decay rate in optomechanical systems, the information can be stored in mechanical modes for a long time.  Hence, entangled optomechanical systems could be profitably used for the
realization of quantum communication networks, in which the mechanical modes
play the role of local nodes where quantum information can be stored and
retrieved, and optical modes transfer this information between the nodes \cite{Mancini2003,Pirandola2003,Pirandola2004,Pirandola2006}.

In fact, any quantum communication application involves traveling output
modes rather than intracavity ones. Therefore, it is very important to study how
to obtain optimal output entanglement in cavity optomechanics.
Here, we use a three-mode optomechanical system (see Fig. 1) to generate output entanglement between two filtered output optical fields. This setup has been
realized in several recent experiments \cite{Dong2012,Hill2012,Andrews2014,Barzanjeh2019}.
Because in such a system the parametric-amplifier interaction and the
beam-splitter interaction can entangle the two intracavity modes (the intracvity entanglement in this model has been  elaborately studied in \cite{Wang2013}), the output
cavity modes are also entangled with each other. In previous theoretical works \cite{Tian2013,Wang2015,Deng2015,Deng2016,Yan2017},
the output entanglement in this model has been studied. But in most of them, the output
entanglement were studied with equal-coupling, small filter bandwidth or without time delay between the two output fields. Especially in \cite{Wang2015}, the authors detailedly studied the output entanglement in the special case of zero bandwidth of filter function.
While large bandwidth is very important for output entanglement because it corresponds to a narrow output field in time domain according to Fourier transform theory. It means that output entanglement between the two filtered fields can be achieved in a short time in the case of large bandwidth.

In this paper, we mainly focus on how to obtain optimal
output entanglement between two filtered optical fields with optimal coupling and optimal time delay for large bandwidth.
First, we obtain the key analytic expression of optimal time delay between the two filtered output fields. We find the time delay will significantly affect the output entanglement in the case of large bandwidth, and we give the reasonable boundary between small bandwidth and large bandwidth.
With large bandwidth and no time delay, the optimal output entanglement will appear at the point where its optimal time delay equals zero, from which we obtain the analytic expression of optimal coupling. Moreover, our
linearized analysis predicts that the entanglement saturates to an optimal value if the optomechanical coupling
is strong enough, and the expression of saturation value is also obtained. Finally, we obtain the optimal output entanglement with optimal time delay and optimal couplings, and find out the output entanglement at resonant frequency is just the optimal one in the whole center frequency domain of output fields.
We believe the results
of this paper are very important to experimental and theoretical physicists
who work on entanglement in cavity optomechanics or circuit quantum electrodynamics.

\begin{figure}[tbp]
\centering\includegraphics[width=7cm]{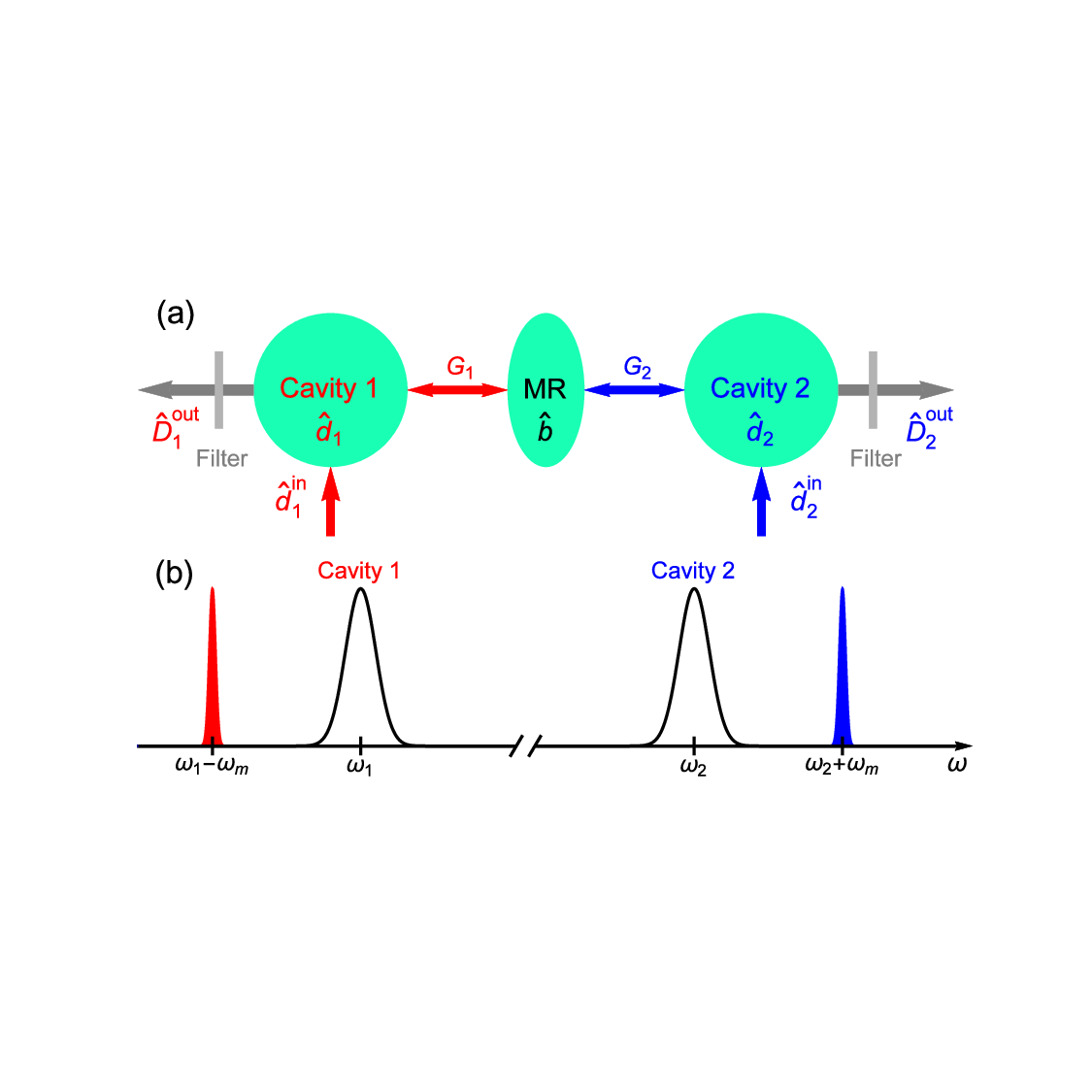}
\caption{ (a) A
three-mode optomechanical system with a mechanical resonator (MR) interacted
with two cavities. Cavity 1 is driven by a red-detuned laser, while cavity 2
is driven by a blue-detuned laser. The entanglement between the filtered
output fields of two cavities $\hat{D}^{\text{out}}_{1}$ and $\hat
{D}^{\text{out}}_{2}$ can be generated. (b) Spectral position of cavity
resonances $\omega_{1}$, $\omega_{2}$ and driving frequencies ($\omega_{1}-\omega_{m}$ and $\omega_{2}+\omega_{m}$).}
\label{Fig1}
\end{figure}

\section{System model}

We consider a three-mode optomechanical system
in which two cavities are coupled to a common mechanical resonator (see Fig. 1).
The Hamiltonian of the system reads
\begin{align}
H  &  =\omega_{\textrm{m}}\hat{b}^{\dag}\hat{b}+\sum_{i=1,2}[\omega_{i}\hat{a}%
_{i}^{\dag}\hat{a}_{i}+g_{i}(\hat{b}^{\dag}+\hat{b})\hat{a}_{i}^{\dag}\hat
{a}_{i}]. \label{Eq1}%
\end{align}
Here, $\hat{a}_{i}$ is the annihilation
operator for cavity $i$ with frequency $\omega_{i}$ and damping rate
$\kappa_{i}$, $\hat{b}$ is the annihilation operator for mechanics resonator
with frequency $\omega_{\textrm{m}}$ and damping rate $\gamma$, and $g_{i}$ is the
optomechanical coupling strength. In order to generate the steady entanglement
between the two output fields, we drive cavity 1 (2) at the red (blue)
sideband with respect to mechanical resonator: $\omega_{d1}=\omega_{1}%
-\omega_{\textrm{m}}$ and $\omega_{d2}=\omega_{2}+\omega_{\textrm{m}}$. If we work in a rotating
frame with respect to the free Hamiltonian, following the standard
linearization procedure, and making the rotating-wave approximation, hence, the Hamiltonian of the system can be written as
\begin{align}
\hat{H}_{\textrm{int}}  &  =G_{1}\hat{b}^{\dag}\hat{d}_{1}+G_{2}\hat{b}\hat{d}_{2}+\textrm{H.c.}
\end{align}
Here, $\hat{d}_{i}=\hat{a}_{i}-\bar{a}_{i}$, $\bar{a}_{i}$ being the classical
cavity amplitude. $G_{i}$ is the effective coupling strength which can be
easily controlled by adjusting the strength of driving fields.
Here, we take $G_{i}$ as real number without loss of generality, and set $G_{i}\ll\omega_{m}$ to satisfy the rotating-wave approximation.

In this paper, we assume that there are no intrinsic photon losses, the cavity damping rates $\kappa_{i}$ are the external coupling rates.
Based on Eq. (2), the dynamics of the system is described by the following
quantum Langevin equations for relevant operators of mechanical and optical
modes%
\begin{align}
\frac{d}{dt}\hat{b}  &  =-\frac{\gamma}{2}\hat{b}-i(G_{1}\hat{d}_{1}+G_{2}%
\hat{d}_{2}^{\dag})-\sqrt{\gamma}\hat{b}^{\textrm{in}},\nonumber\\
\frac{d}{dt}\hat{d}_{1}  &  =-\frac{\kappa_{1}}{2}\hat{d}_{1}-iG_{1}\hat
{b}-\sqrt{\kappa_{1}}\hat{d}_{1}^{\textrm{in}},\label{Eq3}\\
\frac{d}{dt}\hat{d}_{2}^{\dag}  &  =-\frac{\kappa_{2}}{2}\hat{d}_{2}^{\dag
}+iG_{2}\hat{b}-\sqrt{\kappa_{2}}\hat{d}_{2}^{\textrm{in},\dag}.\nonumber
\end{align}
Here, $\hat{b}^{\textrm{in}},\hat{d}_{i}^{\textrm{in}}$
are the input noise operators of mechanical resonator and cavity $i (i=1,2)$,
whose correlation functions are $\langle\hat{b}^{\textrm{in},\dag}(t)\hat{b}%
^{\textrm{in}}(t^{\prime})\rangle=N_{\textrm{m}}\delta(t-t^{\prime})$ and $\langle\hat{d}%
_{i}^{\textrm{in},\dag}(t)\hat{d}_{i}^{\textrm{in}}(t^{\prime})\rangle=N_{i}\delta(t-t^{\prime
})$ respectively. $N_{\textrm{m}}$ and $N_{i}$ are the average thermal populations of
mechanical mode and cavity $i$, respectively. In this paper, we assume these average thermal populations are zero for simplicity, and focus on the interesting regime of strong cooperativities
$C_{i}\equiv4G_{i}^{2}/(\gamma\kappa_{i})\gg1$ and $\kappa_{i}\gg\gamma$. According to the Routh-Hurwitz stability conditions
\cite{DeJesus1987}, the stability condition of our system can be obtained as
$G_{1}^{2}/G_{2}^{2}>\max(\kappa_{1}/\kappa_{2},\kappa_{2}/\kappa_{1})$ for
$\kappa_{1}\neq\kappa_{2}$, and the system is always stable if $\kappa
_{1}=\kappa_{2}$ and $G_{2}\leq G_{1}$ \cite{Wang2013,Wang2015}.

\section{Output fields and optimal time delay}

In a quantum network,
entangled photon pairs are a useful resource for quantum information
processing. Here, the output entanglement can be generated when the output
optical fields pass through the filter functions $f(\omega)$ which satisfy
$\int_{-\infty}^{+\infty}|f(\omega)|^{2}d\omega=1$. From a continuous output field one can extract many independent optical modes, by selecting different time intervals or, equivalently, different frequency intervals \cite{Genes2008}. Here, we adopt a
rectangle filter function with a bandwidth $\sigma$ centered about the frequency
$\omega$ to generate the output temporal modes, i.e., $f(\omega^{\prime
})=\{\theta[\omega^{\prime}-(\omega-\frac{\sigma}{2})]-\theta
[\omega^{\prime}-(\omega+\frac{\sigma}{2})]\}/\sqrt{\sigma}$ with
$\theta[\omega]$ the Heaviside step function. Then, the filtered optical
output fields can be written as%
\begin{equation}
\hat{D}_{i}^{\textrm{out}}[\omega,\sigma,\tau_{i}]=\frac{1}{\sqrt{\sigma}}\int%
_{\omega-\frac{\sigma}{2}}^{\omega+\frac{\sigma}{2}}d\omega^{\prime}e^{-i\omega^{\prime}\tau_{i}}\hat
{d}_{i}^{\textrm{out}}(\omega^{\prime}).
\end{equation}
Here, $\tau_{i}$ is the absolute
time at which the wave packet of interest is emitted from cavity $i$, and $\hat
{d}_{i}^{\textrm{out}}(\omega^{\prime})$ is the Fourier transformation of the output
operators $\hat
{d}_{i}^{\textrm{out}}(t)$ which can be obtained as $\hat{d}_{i}^{\textrm{out}}(t)=\sqrt{\kappa_{i}}\hat{d}_{i}(t)+\hat{d}^{in}_{i}(t)$ according to the input-output relation \cite{Gardiner2004}. The time delay between this two output fields is defined as
$\tau=\tau_{1}-\tau_{2}$. Without
loss of generality, we set $\tau_{1}=\tau$ and $\tau_{2}=0$. The covariance matrix of the
output operators can be computed via the scattering matrix method \cite{Wang2015}. Then, we can use the logarithmic
negativity \cite{Vidal2002,Plenio2005} to quantify the entanglement between
the filtered output cavity modes $\hat{D}_{1}^{\textrm{out}}[\omega,\sigma,\tau]$ and
$\hat{D}_{2}^{\textrm{out}}[-\omega,\sigma,0]$. For simplicity, we write $\hat{D}_{i}^{\textrm{out}}%
[\omega,\sigma,\tau_{i}]$ as $\hat{D}_{i}$ and set equal
cavity damping rate $\kappa_{1}=\kappa_{2}=\kappa$ in the following.

It can be verified that only three correlators $\langle\hat{D}^{\dag}_{1}\hat
{D}_{1}\rangle$, $\langle\hat{D}^{\dag}_{2}\hat
{D}_{2}\rangle$ and $\langle\hat{D}_{1}\hat
{D}_{2}\rangle$ are nonzero in our system and the two-mode squeezed thermal state $\hat{\rho}_{12}=\hat{S}_{12}(R_{12})[\hat{\rho}_{1}^{\textrm{th}}(\bar{n}_{1})\otimes\hat{\rho}_{2}^{\textrm{th}}(\bar{n}_{2})]\hat{S}_{12}^{\dag}(R_{12})$ with $\hat{S}_{12}(R_{12})=\exp[R_{12}\hat{D}_{1}\hat{D}_{2}-\textrm{H.c.}]$. Here, $\hat{S}_{12}$ is the two-mode squeeze operator with $R_{12}$ being the squeezing parameter, and $\hat{\rho}_{i}^{\textrm{th}}(\bar{n}_{i})$ describes a single-mode thermal state with
average population $\bar{n}_{i}$. Thus, the covariance matrix of the system is the same as the two-mode squeezed thermal state. Since two Gaussian states with the same covariance matrix represent the same state, the output cavity state of our system can be mapped to the two-mode squeezed thermal state. The concrete relationship between the two-mode squeezed thermal
state and our system can be obtained as follows: $2\bar{n}_{1}=\langle\hat{D}^{\dag}_{1}\hat
{D}_{1}\rangle-\langle\hat{D}^{\dag}_{2}\hat
{D}_{2}\rangle-1+\sqrt{A^{2}-4|\langle\hat{D}_{1}\hat
{D}_{2}\rangle|^{2}}$, $2\bar{n}_{2}=\langle\hat{D}^{\dag}_{2}\hat
{D}_{2}\rangle-\langle\hat{D}^{\dag}_{1}\hat
{D}_{1}\rangle-1+\sqrt{A^{2}-4|\langle\hat{D}_{1}\hat
{D}_{2}\rangle|^{2}}$, and $\tanh{2R_{12}}=2|\langle\hat{D}_{1}\hat
{D}_{2}\rangle|/A$ with $A=\langle\hat{D}^{\dag}_{1}\hat
{D}_{1}\rangle+\langle\hat{D}^{\dag}_{2}\hat
{D}_{2}\rangle+1$. From these relations, it can be concluded that the optimal time delay for entanglement between the two filtered output fields is the time delay that makes the modulus $|\langle\hat{D}_{1}\hat{D}_{2}\rangle|$ reach a maximum. Using the input-output relation \cite{Gardiner2004} and scattering matrix
method, we obtain the expression of correlator $\langle\hat{D}_{1}\hat{D}_{2}\rangle$ as
\begin{eqnarray}
\langle\hat{D}_{1}\hat
{D}_{2}\rangle=\int_{\omega-\frac{\sigma}{2}}^{\omega+\frac{\sigma}{2}}\frac{4G_{1}G_{2}\kappa[2G^{2}_{1}(\kappa-2i\Omega)+(\kappa+2i\Omega)(2G^{2}_{2}+2\Omega^{2}+i\Omega\kappa)]e^{-i \Omega\tau}}{\sigma(\kappa^{2}+4\Omega^{2})[(8G^{2}_{1}-8G^{2}_{2}-\kappa^{2})\Omega^{2}-4(G^{2}_{1}-G^{2}_{2})^{2}-4\Omega^{4}]}d\Omega.
\end{eqnarray}
Here, we assume the mechanical damping rate $\gamma$ is much less than other parameters ($\kappa, G_{1}, G_{2}$), which is reasonable in most cases in cavity optomechanics.
It can be seen from Eq. (5) that the time delay $\tau$ will significantly affect the modulus of the correlator $\langle\hat{D}_{1}\hat
{D}_{2}\rangle$ for large bandwidth, while it almost has no effect on the modulus for small bandwidth because now the factor $e^{-i\Omega\tau }$ can be extracted out of the integration in the limit of $\sigma\rightarrow0$. It is hard to analytically calculate Eq. (5) with large bandwidth $\sigma$ due to the factor $e^{-i\Omega\tau}$ in the integration. We find that if the time delay is much less than the width of wave-packet in time domain, i.e., $\tau \ll\sigma^{-1}$, we can expand the factor of $e^{-i\Omega\tau}$ by Taylor's series to quadratic terms of $\tau$ (here, we study the case of resonant frequency (i.e., $\omega=0$ in the rotating frame), hence we have $\Omega\tau<\sigma\tau\ll1$). Substitute the expanded terms to Eq. (5) and integrate, we can obtain the approximate analytical expression of $\langle\hat{D}_{1}\hat
{D}_{2}\rangle$. Then the optimal time delay $\tau_{\text{opt}}$ can be obtained by solving the equation $\frac{\partial|\langle\hat{D}_{1}\hat
{D}_{2}\rangle|}{\partial\tau}=0$. Through complex calculation, we obtain the key analytical expression of $\tau_{\text{opt}}$ for resonant frequency as follows
\begin{equation}
\tau_{\text{opt}}=\frac{20(G_{2}^{2}-G_{1}^{2})+5\kappa^{2}+3\sigma^{2}%
}{10(G_{1}^{2}+G_{2}^{2})\kappa}.
\end{equation}
For the special case of equal coupling ($G_{1}=G_{2}$) and $\sigma\ll\kappa$, the expression of
$\tau_{\text{opt}}$ will become $\kappa/4G^{2}_{1}$ which is consistent with
the result in \cite{Wang2015,Deng2016}. But now the optimal time delay Eq. (6) is a more general expression, i.e., it is still valid even for very large bandwidth (e.g. $\sigma=\kappa$, see the next). It can be seen from Eq. (6) the optimal time delay $\tau_{\text{opt}}$ will evolve
from negative to positive with the
increase of $G_{2}$ while keeping
other three parameters unchanged. We will see that the coupling $G_{2}$ at which the
optimal time delay equals zero is a very special coupling. In the following, we will study the output entanglement in two cases, without and with time delay respectively.

\section{Optimal output entanglement without time delay}

\begin{figure}[b]
\centering\includegraphics[width=7cm]{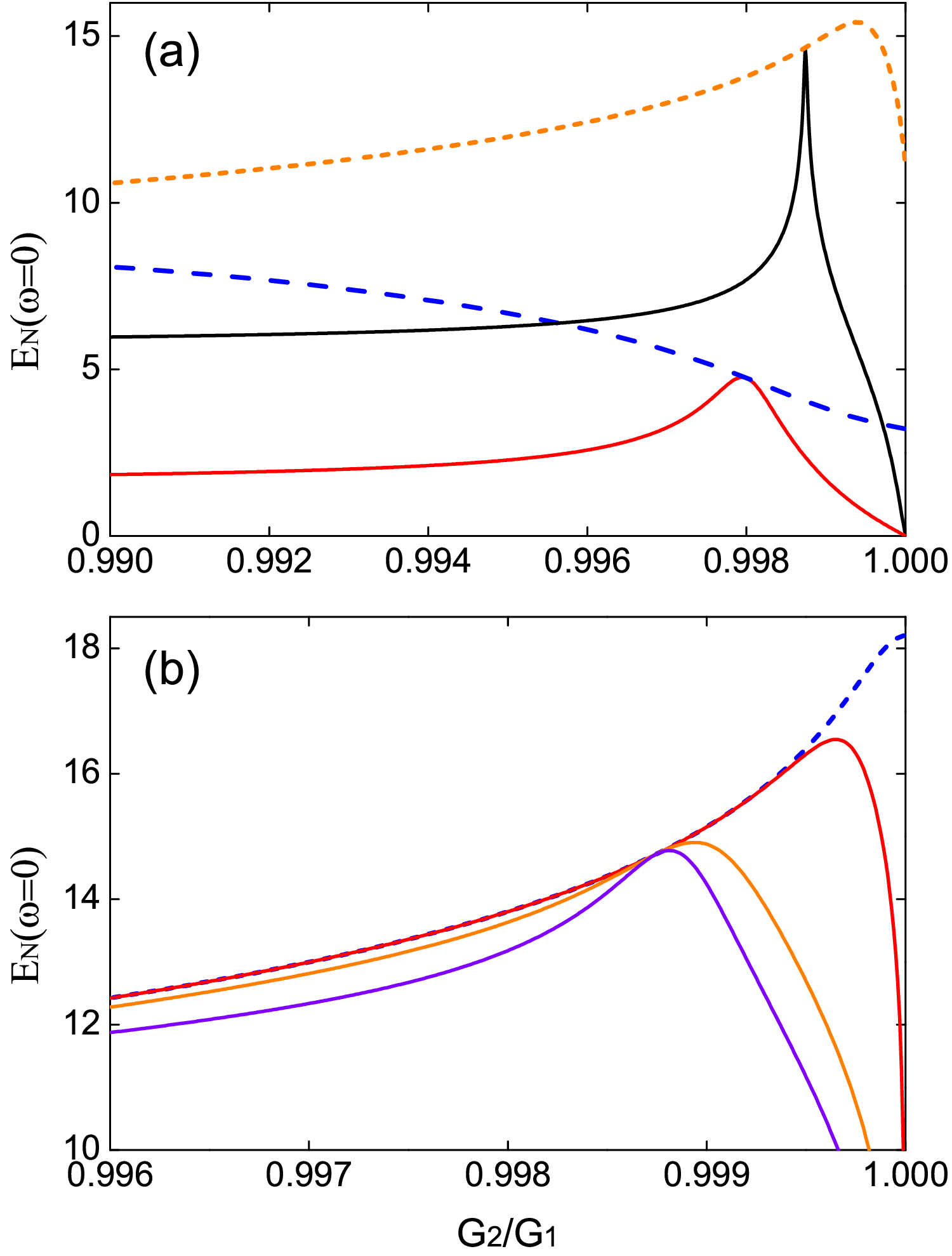}
\caption{ (a)
The output entanglement $\textrm{E}_{\textrm{N}}(\omega=0)$ are plotted vs $G_{2}/G_{1}$ with large bandwidth $\sigma=\kappa/10$ (black line), $\sigma=\kappa$ (red line), and the two dashed lines are their corresponding ones with numerical optimal
time delay. (b) The output entanglement $\textrm{E}_{\textrm{N}}(\omega=0)$ are plotted vs $G_{2}/G_{1}$ with small bandwidth $\sigma=\kappa/10^{4}$ (red line),  $\sigma=2\kappa/10^{3}$ (yellow line), the boundary bandwidth $\sigma_{b}=\frac{\sqrt{3}\kappa^{3}}{4G_{1}^{2}}$ (violet line), and their common line with numerical optimal time delay (blue dashed line). The other parameters are $\gamma=1, \kappa=10^{5}, G_{1}=10\kappa$.}%
\label{Fig2}%
\end{figure}

In this part, we study how to obtain the optimal coupling and the optimal
output entanglement without time delay $(\tau=0)$. The
entangling interaction $G_{2}(\hat{b}\hat{d}_{2}+\hat{b}^{\dag}\hat{d}%
_{2}^{\dag})$ in $\hat{H}_{\text{int}}$ entangles the mechanical resonator
$\hat{b}$ and cavity mode $\hat{d}_{2}$, and the beam splitter interaction
$G_{1}(\hat{b}^{\dag}\hat{d}_{1}+\hat{b}\hat{d}_{1}^{\dag})$ swaps the
$\hat{b}$ and $\hat{d}_{1}$ states, these two combined interactions yield the
net entanglement between $\hat{d}_{1}$ and $\hat{d}_{2}$. The entanglement generated within the cavities can be transferred to the filtered output fields. It is obvious that
the output entanglement will disappear as coupling $G_{2}=0$. Further considering that the
output entanglement $\textrm{E}_{\textrm{N}}(\omega=0)$ will be strongly suppressed with the increasing of filter bandwidth, and approaches zero in the case of $G_{2}=G_{1}$ \cite{Wang2015,Yan2017}, we judge
that there exists an optimal coupling $G^{\text{opt}}_{2}$ (0<$G^{\text{opt}}_{2}$<$G_{1}$) that makes the
output entanglement $\textrm{E}_{\textrm{N}}(\omega=0)$ reach its maximum.

To derive the expression of optimal coupling $\textrm{G}^{\text{opt}}_{2}$, we note that the output entanglement with optimal time delay should be
larger than (equal to) that without time delay. In addition, it can be seen from Eq. (6)
that the optimal time delay may equal zero. It means that the curve of output
entanglement with optimal time delay will be tangent to that without time
delay at the point where the optimal time delay equals zero.
In Fig. 2(a), the output entanglement $\textrm{E}_{\textrm{N}}(\omega=0)$ is plotted vs
$G_{2}/G_{1}$ with large filter bandwidth $\sigma=\kappa$ (red line), $\sigma=\kappa/10$ (black line), and the dashed lines are their corresponding lines with numerical optimal time delay. The other parameters are $\gamma=1, \kappa=10^{5},$ and $G_{1}=10\kappa$.
It can be clearly seen from Fig. 2(a) that the output entanglement takes the maximum at the tangent point for large
bandwidth. Hence, according to Eq. (6) (setting $\tau_{\text{opt}}=0$), the optimal coupling for large
bandwidth can be obtained as
\begin{equation}
\textrm{G}_{2}^{\text{opt}}=\frac{1}{2}\sqrt{4\textrm{G}_{1}^{2}-\kappa^{2}-\frac{3\sigma^{2}}{5}}.
\end{equation}
In Fig. 2(b), we plot the
output entanglement $\textrm{E}_{\textrm{N}}(\omega=0)$ vs
$G_{2}/G_{1}$ for small bandwidth $\sigma=\kappa/10^{4}$ (red line), $\sigma=2\kappa/10^{3}$ (yellow line) and their corresponding common line with numerical optimal time delay (blue dashed line). It can be seen from Fig. 2(b) that the maximum of output entanglement will not appear
at the tangent point. This is because the time
delay has no significantly effect on the output entanglement in the case of small bandwidth. For small bandwidth, the optimal coupling can be obtained as
\begin{equation}
\textrm{G}_{2}^{\text{opt}}=\textrm{G}_{1}+\frac{\textrm{G}_{1}\sigma}{2\sqrt{3}\kappa}-\frac
{\sqrt{\kappa\sigma}}{2\sqrt[4]{3}}.
\end{equation}
From Eqs. (7) and (8), we obtain the boundary between small bandwidth and large
bandwidth as $\sigma_{b}=\frac{\sqrt{3}\kappa^{3}}{4G_{1}^{2}}$.
We also plot the output entanglement $\textrm{E}_{\textrm{N}}(\omega=0)$ with boundary bandwidth $\sigma=\sigma_{b}$ (violet line) in Fig. 2(b) from which it can be seen the maximum value of output entanglement will not appear at the tangent point anymore as $\sigma<\sigma_{b}$.

At the last of this part,
we give an interesting result for large bandwidth, i.e., entanglement saturation. We find the optimal output entanglement $\textrm{E}^{\textrm{opt}}_{\textrm{N}}(\omega=0)$ (with optimal coupling Eq. (7)) will saturate to an optimal value as $\textrm{G}_{1}\gg\sqrt{\frac{\kappa^{5}}{\sqrt{3}\sigma^{3}}}$. In Fig. 3, we plot the optimal
output entanglement $\textrm{E}^{\textrm{opt}}_{\textrm{N}}(\omega=0)$ vs $\textrm{G}_{1}/\kappa$ for bandwidth $\sigma=\kappa$ (red line), $\sigma=\kappa/2$ (blue line), $\sigma=\kappa/10$ (black line). It can be seen clearly from Fig. 3 that the optimal output
entanglement will approach a constant (saturation value) with the increase of coupling $\textrm{G}_{1}$. The saturation value can be obtained as
\begin{equation}
\textrm{E}_{\textrm{N}}^{\text{sat}}=-\ln[\sqrt{\frac{\left(  \kappa^{2}+\sigma^{2}\right)
\left(  15\kappa^{2}\sigma+4\sigma^{3}-3\alpha\beta\right)  ^{2}}{9\kappa
^{2}\sigma^{2}\alpha^{2}}}]
\end{equation}
with $\alpha=5\kappa^{2}+3\sigma^{2}, \beta=\kappa\arctan\left(\frac{\sigma
}{\kappa}\right).$
The saturation value plotted according to Eq. (9) in Fig. 3 (three green dashed lines) perfectly fits the numerical result. Due to entanglement saturation, the coupling $\textrm{G}_{1}$ is not the stronger the better. It can also be seen from Fig. 3 that for larger bandwidth, such as $\sigma=\kappa$ (red line), the coupling $\textrm{G}_{1}$ slightly greater than $\kappa$ can make output entanglement saturation occur. For the case of $\sigma\ll\kappa$, the saturation value Eq. (9) can be simplified as $\ln[\frac{175\kappa^{6}}{4\sigma^{6}}]$. These results are important to experimental physicists working on entanglement in cavity-optomechanics.

\begin{figure}[htbp]
\centering\includegraphics[width=7cm]{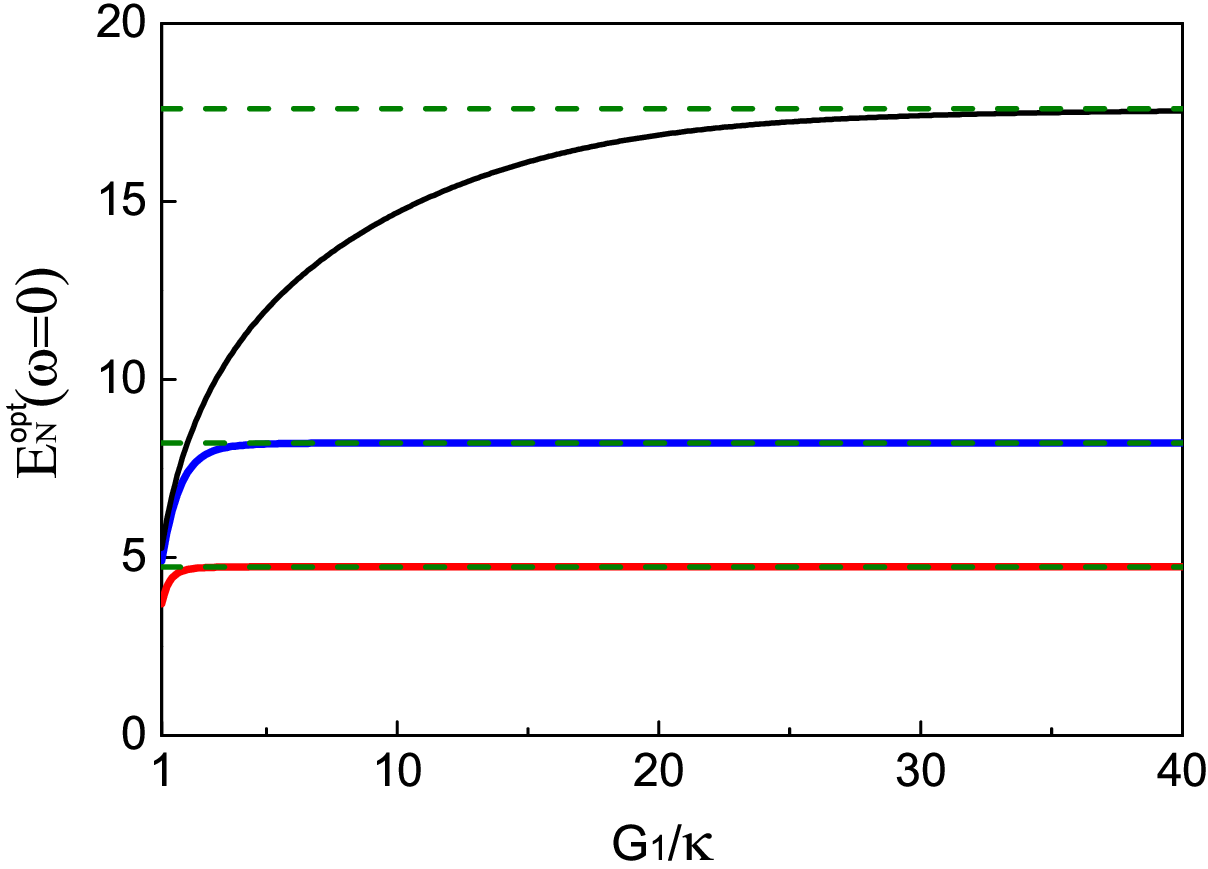}
\caption{ The optimal output entanglement $\textrm{E}^{\textrm{opt}}_{\textrm{N}}(\omega=0)$ vs $G_{1}/\kappa$ with optimal coupling Eq. (7) for bandwidth $\sigma=\kappa/10$ (black line), $\sigma=\kappa/2$ (blue line), $\sigma=\kappa$ (red line), and the saturation values are plotted according to Eq. (9) (green dashed lines). The other parameters are $\gamma=1, \kappa=10^{5}, G_{1}=10\kappa$.}%
\label{Fig3}%
\end{figure}

\section{Optimal output entanglement with time delay}

Now, we study how to
find the optimal output entanglement with optimal time
delay $(\tau=\tau_{\textrm{opt}})$. First, we plot the optimal time delay vs. $G_{2}/G_{1}$ according to Eq. (6) (blue solid line) and the
numerical result (red dashed line) in Fig. 4(a). The optimal time delay will become positive with the increase of $G_{2}$ (see the inset of Fig. 4(a)). In Fig. 4(b), we plot the corresponding output
entanglement with these two optimal time delays with parameters $\gamma=1, \kappa=10^{5}, \sigma=\kappa,
G_{1}=10\kappa$. The two curves of output entanglement fit very well except for a very small area (see the inset of Fig. 4(b)). With these parameters, the maximum of output entanglement appears at optimal coupling $G^{\textrm{opt}}_{2}\approx0.983G_{1}$ (see Fig. 4(b)) rather than
the point $G_{2}\approx0.998G_{1}$ where the optimal time delay equals zero. If $G_{1}\gg\kappa,\sigma$, we can obtain the optimal coupling and the optimal output entanglement as
\begin{equation}
G^{\textrm{opt}}_{2}=G_{1}-\left(  \frac{\alpha\left(  15\kappa^{2}\sigma+4\sigma^{3}-3\alpha
\beta\right)  }{400\left(  6\sigma-3\beta\right)}\right)^{1/4}
\end{equation}
and
\begin{equation}
\textrm{E}_{\textrm{N}}^{\text{opt}}=-\ln[\sqrt{\frac{\alpha\left(  2\sigma-\beta\right)
\left(  15\kappa^{2}\sigma+4\sigma^{3}-3\alpha\beta\right)  }{4800G_{1}^{4}
\sigma^{2}}}].
\end{equation}
We also plot the optimal output entanglement (the highest point) according to Eqs. (10) and (11) in Fig. 4(b) (see the green dot), and it fits the numerical result very well.

\begin{figure}[htbp]
\centering\includegraphics[width=7cm]{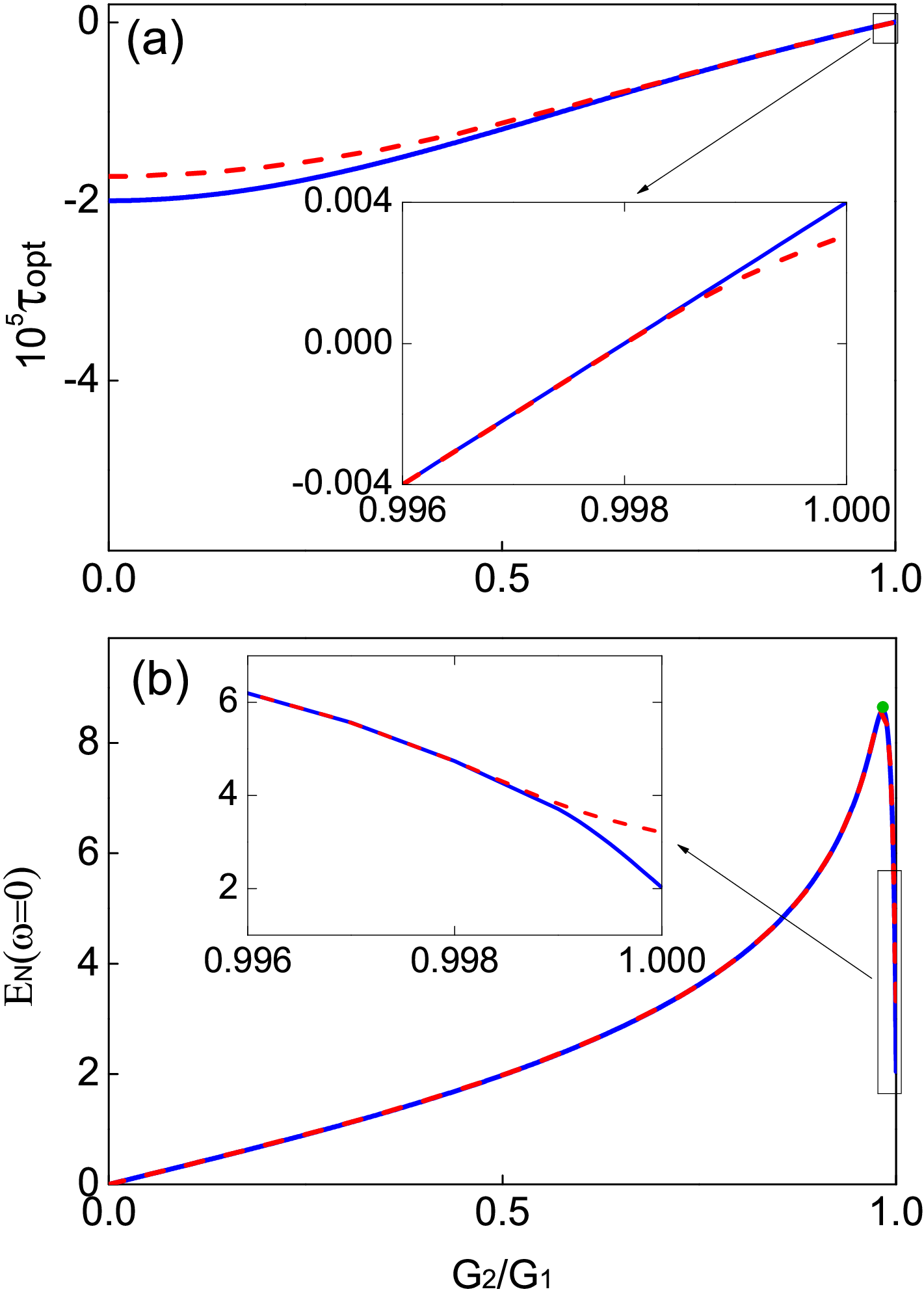}
\caption{ (a)
The optimal time delay $\tau_{\textrm{opt}}$ are plotted vs $G_{2}/G_{1}$ according to Eq. (6) (blue solid line) and the numerical result (red dashed line).
(b) The output entanglement $\textrm{E}_{\textrm{N}}(\omega=0)$ are plotted vs $G_{2}/G_{1}$ with the optimal time delay according to Eq. (6) (blue solid line) and the numerical result (red dashed line). And the optimal output entanglement (see the green dot) is plotted according to Eqs. (10) and (11). The parameters are $\gamma=1, \kappa=10^{5}, \sigma=\kappa,
G_{1}=10\kappa$.}%
\label{Fig4}%
\end{figure}

\begin{figure}[htbp]
\centering\includegraphics[width=7cm]{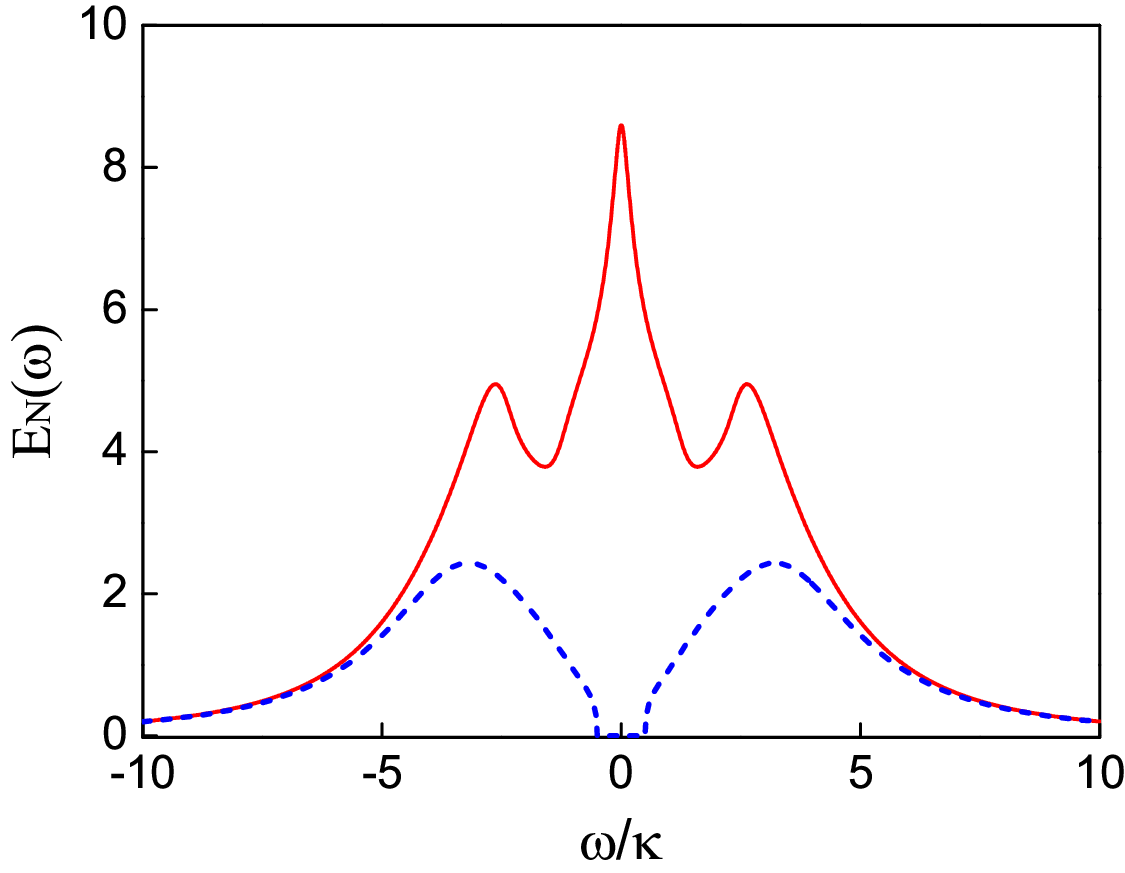}
\caption{ The output entanglement $\textrm{E}_{\textrm{N}}(\omega)$ are plotted vs the normalized center frequency $\omega/\kappa$, with numerical optimal time delay and optimal coupling Eq. (10) (red solid line), with no time delay ($\tau=0$) and equal-coupling ($\textrm{G}_{1}=\textrm{G}_{2}$) (blue dashed line). The parameters are $\gamma=1, \kappa=10^{5}, \sigma=\kappa,
G_{1}=10\kappa$.}%
\label{Fig5}%
\end{figure}

Until now, we study the output entanglement just for resonant frequency $(\omega=0)$. In Fig. 5, we plot output entanglement
$\textrm{E}_{\textrm{N}}(\omega)$ vs. $\omega/\kappa$ with optimal coupling Eq. (10) and with numerical optimal time delay (red line).
It is obviously seen from Fig. 5 that the optimal output entanglement Eq. (11) is just the optimal one in the whole center frequency domain of output fields.
For contrast, we also plot the output entanglement with equal-coupling ($G_{2}=G_{1}$) and without time delay ($\tau=0$) (blue dashed line).
It can be seen from Fig. 5 the output entanglement in the vicinity of resonant frequency can be largely enhanced by adopting optimal coupling Eq. (10) and optimal time delay.
In experiment, it just need to adjust the coupling strength $G_{2}$ according to the parameters of the systems to obtain the optimal output entanglement.

\section{Conclusions}

In summary, we have studied theoretically how to obtained the optimal output entanglement between two filtered
output fields with optimal coupling and optimal time delay in a three-mode cavity optomechanical system. We obtain the key expression of optimal time delay between the two filtered output fields, and give a reasonable boundary between large bandwidth and small bandwidth. Besides, we draw three important
conclusions: (1) with large bandwidth and no time delay, the optimal output entanglement will emerge at the point where the optimal time delay equals zero; (2) our linearized analysis predicts that the optimal output entanglement with large bandwidth will saturate if the coupling is strong enough;
(3) using the optimal coupling Eq. (10) and optimal time delay, we find the optimal output entanglement Eq. (11) for resonant frequency ($\omega=0$) is just the optimal one in the whole center frequency domain of output fields.
Our results can also be applied to other parametrically coupled three-mode
bosonic systems, and may be useful to experimentalists to obtain large entanglement.

\section*{Funding}
National Natural Science Foundation of China (NSFC) (11574398, 11847018, and 11861131001).

%Using the \texttt{cite.sty} package will make these citations appear like so: [2--4].

%%%%%%%%%%%%%%%%%%%%%%% References %%%%%%%%%%%%%%%%%%%%%%%%%

%%%%%%%%%% If using BibTeX:
%\bibliography{sample}

\begin{thebibliography}{1}
\newcommand{\enquote}[1]{``#1''}


\bibitem{Samuel2005} S. L. Braunstein, P. van Loock, \enquote{Quantum information with continuous variables,} Rev. Mod. Phys. \textbf{77}, 513--577 (2005).
\bibitem {Weedbrook2012} C. Weedbrook, S. Pirandola, R. Garc\'{\i}a-Patr\'on, N. J. Cerf, T. C. Ralph, J. H. Shapiro, and S. Lloyd, \enquote{Gaussian quantum information,} Rev. Mod. Phys. \textbf{84}, 621--669 (2012).

\bibitem {Julsgaard2011}B. Julsgaard, A. Kozhekin, and E. S. Polzik, \enquote{Experimental long-lived entanglement of two macroscopic objects,} Nature
\textbf{413}, 400--403 (2001).

\bibitem {Krauter2011}H. Krauter, C. A. Muschik, K. Jensen, W. Wasilewski, J.
M. Petersen, J. I. Cirac, and E. S. Polzik, \enquote{Entanglement generated by dissipation and steady state entanglement of two macroscopic objects,} Phys. Rev. Lett. \textbf{107},
080503 (2011).

\bibitem {Berkley2003}A. J. Berkley, H. Xu, R. C. Ramos, M. A. Gubrud, F. W.
Strauch, P. R. Johnson, J. R. Anderson, A. J. Dragt, C. J. Lobb, F. C.
Wellstood, \enquote{Entangled macroscopic quantum states in two superconducting qubits,} Science \textbf{300}, 1548--1550 (2003).

\bibitem {Neeley2010}M. Neeley, R. C. Bialczak, M. Lenander, E. Lucero, M.
Mariantoni, D. Sank, H. Wang, M. Weides, J. Wenner, Y. Yin, T. Yamamoto, A. N.
Cleland, and J. M. Martinis, \enquote{Generation of three-qubit entangled states using superconducting phase qubits,} Nature \textbf{467}, 570 (2010).

\bibitem {DiCarlo2010}L. DiCarlo, M. Reed, L. Sun, B. L. Johnson, J. M. Chow,
J. M. Gambetta, L. Frunzio, S. M. Girvin, M. H. Devoret, and R. J. Schoelkopf, \enquote{Preparation and measurement of three-qubit entanglement in a superconducting circuit,}
Nature \textbf{467}, 574 (2010).

\bibitem {Flurin2012}E. Flurin, N. Roch, F. Mallet, M. H. Devoret, and B.
Huard, \enquote{Generating entangled microwave radiation over two transmission lines,} Phys. Rev. Lett. \textbf{109}, 183901 (2012).

\bibitem {Palomaki2013} T. A. Palomaki, J. D. Teufel, R. W. Simmonds, and K. W. Lehnert, \enquote{Entangling mechanical motion with microwave fields,} Science \textbf{342}, 710 (2013).

\bibitem {Mancini2003} S. Mancini, D. Vitali, and P. Tombesi, \enquote{Scheme for teleportation of quantum states onto a mechanical resonator,} Phys. Rev. Lett. \textbf{90}, 137901 (2003).
\bibitem {Pirandola2003} S. Pirandola, S. Mancini, D. Vitali, and P. Tombesi, \enquote{Continuous-variable entanglement and quantum-state teleportation between optical and macroscopic vibrational modes through radiation pressure,} Phys. Rev. A \textbf{68}, 062317 (2003).
\bibitem {Pirandola2004} S. Pirandola, S. Mancini, D. Vitali, and P. Tombesi, \enquote{Light reflection upon a movable mirror as a paradigm for continuous variable teleportation network,} J. Mod. Opt \textbf{51}, 901 (2004).
\bibitem {Pirandola2006} S. Pirandola, D. Vitali, P. Tombesi, and S. Lloyd, \enquote{Macroscopic entanglement by entanglement swapping,} Phys. Rev. Lett. \textbf{97}, 150403 (2006).
\bibitem {Kiesewetter2014} S. Kiesewetter, Q. Y. He, P. D. Drummond, and M. D. Reid, \enquote{Scalable quantum simulation of pulsed entanglement and Einstein-Podolsky-Rosen steering in optomechanics,} Phys. Rev. A \textbf{90}, 043805 (2014).

\bibitem {Bhattacharya2008}M. Bhattacharya, P.-L. Giscard, and P. Meystre, \enquote{Entangling the rovibrational modes of a macroscopic mirror using radiation pressure,}
Phys. Rev. A \textbf{77}, 030303(R) (2008).

\bibitem {Chen2014}R. X. Chen, L. T. Shen, Z. B. Yang, H. Z. Wu, and S. B.
Zheng, \enquote{Enhancement of entanglement in distant mechanical vibrations via modulation in a coupled optomechanical system,} Phys. Rev. A \textbf{89}, 023843 (2014).

\bibitem {Liao2014}J. Q. Liao, Q. Q. Wu, and F. Nori, \enquote{Entangling two macroscopic mechanical mirrors in a two-cavity optomechanical system,} Phys. Rev. A \textbf{89}, 014302 (2014).

\bibitem {Yang2015}C. J. Yang, J. H. An, W. Yang, and Y. Li, \enquote{Generation of stable entanglement between two cavity mirrors by squeezed-reservoir engineering,} Phys. Rev. A
\textbf{92}, 062311 (2015).

\bibitem {Paternostro2007}M. Paternostro, D. Vitali, S. Gigan, M. S. Kim, C.
Brukner, J. Eisert, and M. Aspelmeyer, \enquote{Creating and probing multipartite macroscopic entanglement with light,} Phys. Rev. Lett. \textbf{99}, 250401 (2007).

\bibitem {Wipf2008}C. Wipf, T. Corbitt, Y. Chen, and N. Mavalvala, \enquote{Route to ponderomotive entanglement of light via optically trapped mirrors,} New J.
Phys. \textbf{10}, 095017 (2008).

\bibitem {Genes2008}C. Genes, A. Mari, P. Tombesi, and D. Vitali, \enquote{Robust entanglement of a micromechanical resonator with output optical fields,} Phys. Rev. A
\textbf{78}, 032316 (2008).

\bibitem {Barzanjeh2011}Sh. Barzanjeh, D. Vitali, P. Tombesi, and G. J.
Milburn, \enquote{Entangling optical and microwave cavity modes by means of a nanomechanical resonator,} Phys. Rev. A \textbf{84}, 042342 (2011).

\bibitem {Barzanjeh2012}Sh. Barzanjeh, M. Abdi,G. J.Milburn, P. Tombesi, and
D.Vitali, \enquote{Reversible optical-to-microwave quantum interface,} Phys. Rev. Lett. \textbf{109}, 130503 (2012).

\bibitem {Barzanjeh2013}Sh. Barzanjeh, S. Pirandola, and C. Weedbrook, \enquote{Continuous-variable dense coding by optomechanical cavities,} Phys.
Rev. A \textbf{88}, 042331 (2013).

\bibitem {Wang2013}Y.-D. Wang and A. A. Clerk, \enquote{Reservoir-engineered entanglement in optomechanical systems,} Phys. Rev. Lett. \textbf{110},
253601 (2013).

\bibitem {Kuzyk2013}M. C. Kuzyk, S. J. van Enk, and H. Wang, \enquote{Generating robust optical entanglement in weak-coupling optomechanical systems,} Phys. Rev. A
\textbf{88}, 062341 (2013).

\bibitem {Vitali2007}D. Vitali, S. Gigan, A. Ferreira, H. R. B$\ddot{o}$hm, P.
Tombesi, A. Guerreiro, V. Vedral, A. Zeilinger, and M. Aspelmeyer, \enquote{Optomechanical entanglement between a movable mirror and a cavity field,} Phys. Rev.
Lett. \textbf{98}, 030405 (2007).

\bibitem {Hofer2011}S. G. Hofer, W. Wieczorek, M. Aspelmeyer, and K. Hammerer, \enquote{Quantum entanglement and teleportation in pulsed cavity optomechanics,}
Phys. Rev. A \textbf{84}, 052327 (2011).

\bibitem {Akram2012}U. Akram, W. Munro, K. Nemoto, and G. J. Milburn, \enquote{Photon-phonon entanglement in coupled optomechanical arrays,} Phys.
Rev. A \textbf{86}, 042306 (2012).

\bibitem {Sinha2015}K. Sinha, S. Y. Lin, and B. L. Hu, \enquote{Mirror-field entanglement in a microscopic model for quantum optomechanics,} Phys. Rev. A \textbf{92}, 023852 (2015).

\bibitem{JLi2015}J. Li, I. Moaddel Haghighi, N. Malossi, S. Zippilli, and D. Vitali, \enquote{Generation and detection of large and robust entanglement between two different mechanical resonators in cavity optomechanics,} New J. Phys. \textbf{17}, 103037 (2015).
\bibitem{JLi2017}J. Li, G. Li, S. Zippilli, D. Vitali, and T. Zhang, \enquote{Enhanced entanglement of two different mechanical resonators via coherent feedback,} Phys. Rev. A \textbf{95}, 043819 (2017).
\bibitem{Asjad2016}M. Asjad, P. Tombesi, and D. Vitali, \enquote{Feedback control of two-mode output entanglement and steering in cavity optomechanics,} Phys. Rev. A \textbf{94}, 052312 (2016).

\bibitem {Bing2014}Q. Lin, B. He, R. Ghobadi, and C. Simon, \enquote{Fully quantum approach to optomechanical entanglement,} Phys.Rev.A \textbf{90}, 022309 (2014)

\bibitem {Tian2013}L. Tian, \enquote{Robust photon entanglement via quantum interference in optomechanical interfaces,} Phys. Rev. Lett. \textbf{110}, 233602 (2013).

\bibitem {Deng2015}Z. J. Deng, S. J. M. Habraken, and F. Marquardt, \enquote{Entanglement rate for Gaussian continuous variable beams,} New J.
Phys. \textbf{18}, 063022 (2016).

\bibitem {Deng2016}Z. J. Deng, X. B. Yan, Y. D. Wang, and C. W. Wu, \enquote{Optimizing the output-photon entanglement in multimode optomechanical systems,} Phys. Rev. A \textbf{93}, 033842 (2016).

\bibitem {Wang2015}Y.-D.Wang, S. Chesi, and A. A. Clerk, \enquote{Bipartite and tripartite output entanglement in three-mode optomechanical systems,} Phys. Rev. A \textbf{91}, 013807 (2015).

\bibitem{Li2013} H.-K. Li, X.-X. Ren, Y.-C. Liu, and Y.-F. Xiao, \enquote{Photon-photon interactions in a largely detuned optomechanical cavity,} Phys. Rev. A \textbf{88}, 053850 (2013).

\bibitem{Sun2017}F. X. Sun, D. Mao, Y. T. Dai, Z. Ficek, Q. Y. He, Q. H. Gong, \enquote{Phase control of entanglement and quantum steering in a three-mode optomechanical system,} New J. Phys. \textbf{19}, 123039 (2017).

\bibitem {Yan2017} X. B. Yan, \enquote{Enhanced output entanglement with reservoir engineering,} Phys. Rev. A \textbf{96}, 053831 (2017).


\bibitem {Dong2012}C. Dong, V. Fiore, M. C. Kuzyk, and H. Wang, \enquote{Optomechanical dark mode,} Science
\textbf{338}, 1609 (2012).

\bibitem {Hill2012}J. T. Hill, A. H. Safavi-Naeini, J. Chan, and O. Painter, \enquote{Coherent optical wavelength conversion via cavity optomechanics,}
Nat. Commun. \textbf{3}, 1196 (2012).

\bibitem {Andrews2014}R. W. Andrews, R. W. Peterson, T. P. Purdy, K. Cicak, R. W.
Simmonds, C. A. Regal, and K. W. Lehnert, \enquote{Bidirectional and efficient conversion between microwave and optical light,} Nat. Phys. \textbf{10}, 321 (2014).
\bibitem{Barzanjeh2019}S. Barzanjeh, E. S. Redchenko, M. Peruzzo, M. Wulf, D. P. Lewis, G. Arnold, and J. M. Fink, \enquote{Stationary entangled radiation from micromechanical motion,} Nature \textbf{570}, 
480--483 (2019).

\bibitem {DeJesus1987}E. X. DeJesus and C. Kaufman, \enquote{Routh-Hurwitz criterion in the examination of eigenvalues of a system of nonlinear ordinary differential equations,} Phys. Rev. A \textbf{35},
5288 (1987).

\bibitem {Gardiner2004}C. Gardiner and P. Zoller, \emph{Quantum Noise}, 3rd
ed. (Springer, New York, 2004).

\bibitem {Vidal2002}G. Vidal and R. F. Werner, \enquote{Computable measure of entanglement,} Phys. Rev. A \textbf{65},
032314 (2002).

\bibitem {Plenio2005}M. B. Plenio, \enquote{Logarithmic negativity: A full entanglement monotone that is not convex,} Phys. Rev. Lett. \textbf{95}, 090503 (2005).




\end{thebibliography}

%%%%%%%%%% If preparing manually:

\end{document}